\documentclass[aps,twocolumn]{revtex4-1}
\usepackage{graphics}
\usepackage{dcolumn}
\usepackage{bm}
\usepackage{amsmath}
\usepackage{hyperref}
\hypersetup{colorlinks=true, urlcolor=blue}

\usepackage{newlfont}
\usepackage{amssymb}
\usepackage{amsfonts}
\usepackage{amsmath}
\usepackage{graphicx}
\usepackage{bm}

\def \be{\begin{equation}}
\def \ee{ \end{equation} }

\begin{document}
\renewcommand*{\DefineNamedColor}[4]{%
   \textcolor[named]{#2}{\rule{7mm}{7mm}}\quad
  \texttt{#2}\strut\\}

\definecolor{red}{rgb}{1,0,0}

\title{Disorder overtakes Order in Information Concentration over Quantum Networks}

\author{R. Prabhu, Saurabh Pradhan, Aditi Sen(De), and Ujjwal Sen}

\affiliation{Harish-Chandra Research Institute, Chhatnag Road, Jhunsi, Allahabad 211 019, India}

\begin{abstract}
We consider different classes of quenched disordered quantum \(XY\) spin chains, including quantum \(XY\) spin glass and 
quantum \(XY\) model with a random transverse field, and investigate the behavior of genuine multiparty entanglement in the ground states of these models. We find that there are distinct ranges of the disorder parameter that gives rise to a higher genuine multiparty entanglement than in the corresponding systems without disorder -- an order-from-disorder in genuine multiparty entanglement. Moreover, we show that such a disorder-induced advantage in the genuine multiparty entanglement is \emph{useful} -- it is almost certainly accompanied by a order-from-disorder for a multiport quantum dense coding capacity with the same ground state used as a multiport quantum network. 
\end{abstract}

\maketitle


\section{Introduction}

Quantum mechanical laws can be utilized to enhance the performance of a large spectrum of 
information transmission protocols -- 
remarkable discoveries in the last two decades -- which can potentially revolutionize  communication networks. Examples include classical information 
transfer through a quantum channel (quantum dense coding) \cite{dc},  arbitrary quantum state transmission with the use of two 
bits of classical communication and a shared quantum state (quantum teleportation) \cite{teleportation}, and secret transmission of 
classical information (quantum cryptography or quantum key distribution) \cite{crypto}. 
 These protocols form the main pillars in the development of quantum communication \cite{physnews}, and at the same time, 
are the basic ingredients for quantum computational tasks \cite{lalholobhor}. That the protocols have already been experimentally 
realized in a variety of substrates, makes the field all the more exciting. 
These include the realizations in photonic systems, by which quantum communication protocols can now be 
implemented between nearby cities and islands (see e.g. \cite{GisinRMP, Zeilinger}, and 
references therein). 
And in ion trap systems, where quantum communication protocols can potentially be useful in a future quantum computer realized in that system 
(see \cite{ionexp, Monroe, Sougato-ion, Hotta, Schmidtkaler,Wineland}, and references therein). 

However, while quantum communication protocols between two parties have received a lot of attention, both on the experimental \cite{GisinRMP, Zeilinger,
ionexp, Monroe, Haroche, abar1, abar2, abar3} and theoretical \cite{HHHHRMP}
fronts, the same is rather limited in the multiparty scenario. This is despite the fact that multi-access quantum communication networks 
involving several senders and receivers have immense applications and form one of the ultimate goals of such studies.

The multiparty scenario presents a very rich structure indeed. See e.g. \cite{multi1, multi2,multi3,multi4,multi5,multi6}, and references therein.
In particular, while capacities for classical as well as quantum information transmission has a one-to-one correspondence with the shared entanglement between the sender and 
the receiver in a bipartite situation (single sender and a single receiver), the multiparty case does 
not lend itself to such a direct correspondence in every setting. 
Certain multi-access quantum capacities of multiparty quantum states can \emph{not} be 
related to its multiparty entanglement content, irrespective of the mode of quantification of the multiparty entanglement \cite{amaderPRA}. 
Here we show that in physically realizable quantum spin models used as  multi-access quantum networks, 
the capacity of transmitting classical information 
is related to the genuine multiparty entanglement content of that network. 
The genuine multiparty entanglement content is quantified by the recently introduced generalized geometric measure \cite{amaderPRA, amaderbound}, 
while the quantum spin models considered are 
the one-dimensional anisotropic quantum \(XY\) models, with nearest-neighbor interactions, and with a transverse field \cite{Sachdev}. 
We consider both ordered as well as 
quenched disordered \(XY\) spin systems. The disordered models that we consider are
(i) the quantum \(XY\) spin glass -- the randomness appearing in the coupling constant, 
(ii) the quantum \(XY\) model with a random transverse field, and 
(iii) the quantum \(XY\) spin glass with a random transverse field. 
Disordered systems form one of the centerstages of studies in many-body physics \cite{eibar-eibar-khukumoni-uthbe, Igloi2005, Misguich2005}. 
In particular, a lot of attention is recently being given to 
quantum spin glass systems (disordered couplings, with or without a disordered transverse field) (see \cite{qsg1, qsg2,qsg3,qsg4,qsg5,Bikas-da}, and references 
therein).


Disordered systems appear due to 
certain environmental conditions and other parameters in the system that are not possible to control experimentally. 
One may expect that disorder would reduce properties like magnetization,
 conductivity, etc., of a given system, and is indeed true for a large variety of systems. However, phenomena where 
disorder enhances magnetization and other properties of a system are also known to occur in both classical and quantum systems 
\cite{Misguich2005, order-from-dis}. 
Studies in quantum transport have shown that disordered systems are always worse than ordered ones for carrying quantum information 
\cite{Sougato1, Sougato2, Sougato3, Sougato4, Sougato5, Sougato6, Sougato7}. 
However, we find that the ground states of a class of quantum spin models with disorder, for certain values of the impurity parameters, can give a clear 
advantage for carrying classical information encoded in 
a quantum state, over the corresponding ordered system. 
Moreover, we 
observe that the disorder can enhance the amount of 
genuine multipartite entanglement of the ground state in the quantum spin  model, and this is found in almost 
the same range in which the disorder-induced amplification of the multiport capacity is obtained. 

Technological limitations currently restricts the physically realizable genuinely multiparty entangled quantum states to about 10 qubits 
\cite{ionexp, Monroe, Schmidtkaler, photonexp, Zeilinger, 14qubit}. Moreover, photonic systems are widely envisaged as the substrates for 
quantum communication protocols, and in that case, currently available technology limitations are even stricter on the number of qubits \cite{4photon,56photon}.
We have therefore focused our attention on quantum communication protocols involving 7 or 8 parties , constituting the sending and receiving 
ports of the communication protocol. However, our investigations with a few more parties have shown that the qualitative behavior remains the same. See the discussion 
in Sec. \ref{conclusion}.


The quantum $XY$ spin chain can be solved exactly by using the  
Jordan-Wigner transformation \cite{Barouch1075}. 
Since we will be interested in finding global properties of the 
multiparty quantum states generated from the corresponding disordered spin models, and in particular will be engaged in finding out a 
genuine multiparty entanglement measure, we will have to carry out the computations by exact diagonalization.

The paper is organized as follows. In Sec. \ref{ordxy}, we briefly describe the one dimensional quantum $XY$ model, without disorder, as well as 
with disorder in the couplings and/or the fields. The disorders that we consider in this paper are of the ``quenched'' type, requiring a ``quenched
average'' of the physical quantities considered in such systems. The meanings of these terms are also briefly explained in Sec. \ref{ordxy}. 
In Sec. \ref{sec:ggm}, we introduce the genuine multiparty entanglement 
measure called generalized geometric measure (GGM) which is used to investigate the multiparty entanglement in the models considered in this paper. 
In Sec. \ref{sec:cap}, 
we describe the classical information transmission protocols that we consider -- along with their capacities --
 for the case of a single sender and a single receiver and as well as for 
multiple senders 
and a single receiver. 
The results are presented in Sec. \ref{sec:ggmcap}. We begin there with the comparison between the \(XY\) spin glass and the ordered \(XY\) systems 
(Sec. \ref{ekta-batrish-1}). 
We find that there is large range of parameters for which the quenched averaged genuine multiparty entanglement (as quantified by GGM) in the 
disordered system
is better than the same in the ordered system -- an order-from-disorder phenomenon 
for a genuine multiparty entanglement. And often there exists a critical coupling constant below which the system shows this 
behavior. A similar picture appears for the multiport channel capacity that we consider. We observe that 
an order-from-disorder for GGM is a prerequisite for the same feature to appear for the multiport capacity. Disorder in the spin glass system is 
introduced through randomness in the interactions. A disordered field term (while maintaining an ordered interaction) is considered in Sec. \ref{ekta-batrish-2}.
Again, order-from-disorder is observed for both GGM and the capacity, but there appeared qualitative differences with the spin glass case. 
It is aproiri unclear as to whether the two types of disorder, if appearing in the same system, would also allow for the order-from-disorder phenomenon. But we 
find that such a situation is indeed allowed, in Sec. \ref{ekta-batrish-3}. We conclude with a discussion in Sec. \ref{conclusion}.

\section{Ordered and Disordered Quantum $XY$ Models}
\label{ordxy}

The one dimensional quantum $XY$ model with nearest neighbor interactions  in a transverse field  is described by  the Hamiltonian  \cite{Sachdev}
\be
\label{XY}
H  = \frac{J}{4}\displaystyle\sum_{\langle ij\rangle}[(1 + \gamma) \sigma_i^x\sigma_{j}^x +
(1-\gamma)\sigma_i^y\sigma_{j}^y]- \frac{h}{2} \displaystyle\sum_i\sigma_i^z,
\ee
where $J$ is the coupling constant and will be positive for antiferromagnetic systems and negative for ferromagnetic systems. 
$\gamma \ne 0$ is the anisotropy constant. $\sigma^x,\, \sigma^y$, and $\sigma^z$ are the Pauli spin matrices
at the corresponding sites,
and $\langle ij\rangle\) \((i,j=1,2,3,\ldots,N)$ indicates that the 
interactions are between all nearest-neighbor spins. 
\(N\) is the length of the spin chain.
Periodic boundary conditions, i.e., $\vec \sigma_{N+1}=\vec \sigma_{N}$ is considered here. Moreover, we assume that \(J,h >0\). 
Such systems can be realized in different physical system, including in ultracold gases \cite{Sachdev, amader-adp}.


Disorder can be introduced in the quantum $XY$ Hamiltonian through several routes.
Here  we consider the following three options: 
(i) the one-dimensional quantum $XY$ spin glass,
where the coupling constant at each site is an independent but  identically distributed random variable,
while the field strength is a constant throughout the chain, 
(ii) the one-dimensional quantum $XY$ model with a random transverse magnetic field,
where the randomness behavior  of the coupling constant and field are interchanged with respect the spin glass system,
and (iii) the one-dimensional quantum $XY$ spin glass with a random transverse magnetic field, where the coupling constant and the field are both 
random. 


\subsection{Quenched disorder and quenched average}

We assume that the disorder in all the different systems considered here are ``quenched''. That is, the time scales in which the 
dynamics of the system takes place is much shorter in 
comparison to the equilibrating times of the disorder. On time scales over which the dynamics of the system takes place, a particular realization of the 
random disorder parameters remains static (``quenched''). 

This has the implication that to find the average of a particular physical quantity, 
the averaging over the disorder has to be performed \emph{after} the calculation of the physical 
parameters for particular realizations of the disorder. 

In the opposite extreme, where the time scales of the dynamics are much longer than the disorder equilibrating times, the disorder is said to be ``annealed'', and 
the corresponding annealed averaging is typically easier to handle.

\subsection{Gaussian random variable}

Along with being quenched, we also assume that the disorder in the models considered, arise from \emph{Gaussian distributed} random variables. 
A random variable \(X\) is said to follow the Gaussian distribution \(N({\cal M}, {\cal S})\), with mean \({\cal M}\) and standard deviation \({\cal S}\), if
the probability density \(P[X=x]\) is 
given by 
\be
\label{gd}
P[X=x]=\frac{1}{{\cal S}\sqrt{2\pi}}\text {exp}\left[-\frac{1}{2}\left(\frac{x-{\cal M}}{{\cal S}}\right)^2\right].
\ee

Typically, the qualitative behavior of an averaged physical quantity in a disordered system is independent of the particular form of the distribution function 
of the random variables involved, and essentially depends on a few first moments of the distribution.

\subsection{The disordered systems and their Hamiltonians}

We now explicitly write down the Hamiltonians of the disordered systems that we consider in this paper. 

\subsubsection{Quantum $XY$ spin glass}


The Hamiltonian of this model is given by \cite{Sachdev} (see Ref. \cite{expt-spin-glass} for experiments)
\be
\label{jihamil}
H_{SG}=\sum_{i}^N \frac{J_i}{4}[(1 + \gamma)\sigma_i^x\sigma_{i + 1}^x+(1 - \gamma)\sigma_i^y\sigma_{i + 1}^y]-\frac{h}{2}\sum_{i}^N\sigma_i^z,
\ee
where the \(J_i\) are independent and identically distributed (i.i.d.) Gaussian quenched random variables, each following  the Gaussian (normal) distribution \(N(J,h)\). 
We assume that \(h>0\).
Note that both the mean and the standard deviation have the dimensions of energy.




\subsubsection{Quantum $XY$ model with random transverse magnetic field}

Reversing the randomness behavior of 
the coupling constant and the field with respect to the spin glass Hamiltonian
leads to the quantum \(XY\) model with a random  transverse field (see \cite{Sachdev, Armand} and references therein),
with the Hamiltonian
\be
\label{hihamil}
H^{RF}=\frac{J}{4}\sum_{i}^N  [(1 + \gamma)\sigma_i^x\sigma_{i + 1}^x+(1 - \gamma)\sigma_i^y\sigma_{i + 1}^y]-\sum_{i}^N\frac{h_i}{2}\sigma_i^z.
\ee
Here \(h_i\) are i.i.d. Gaussian quenched random variables, each following the Gaussian distribution \(N(h,J)\), and \(J>0\).


\subsubsection{Quantum $XY$ spin glass with random transverse magnetic field}

Finally, we consider the case of the 
one-dimensional quantum $XY$ spin glass with random transverse magnetic fields $h_i$ applied in the $z$-direction 
(see \cite{Sachdev, Fischer-etc, ZanardiPRL, ZanardiPRA, Igloi2005, Griffiths-original}
and references therein).
The Hamiltonian for this model is given by
\be
\label{jihihamil}
H^{RF}_{SG}=\sum_{i}^N \frac{J_i}{4} [(1 + \gamma)\sigma_i^x\sigma_{i + 1}^x+(1 - \gamma)\sigma_i^y\sigma_{i + 1}^y]-\sum_{i}^N\frac{h_i}{2}\sigma_i^z,
\ee
where the $J_i$ and $h_i$ are all i.i.d. quenched random variables, with the \(J_i\) being Gaussian distributed as \(N(J,\kappa)\), and the 
\(h_i\) as \(N(h,\kappa)\). Here \(\kappa\) is a positive quantity having the dimensions of energy.




\section{Genuine Multipartite Entanglement Measure}
\label{sec:ggm}

In this section, we introduce the genuine multipartite entanglement measure that will be used in this paper to 
 investigate the behavior of multiparty entanglement in the systems under study. The multipartite entanglement measure  that we consider here 
was introduced in Refs. \cite{amaderPRA, amaderbound}, and is called the generalized geometric measure. For a given \(N\)-party 
pure quantum state \(|\psi_N\rangle \), 
the GGM  is defined as 
\begin{equation}
{\cal E}(\psi_N) = 1 - \mbox{max} |\langle \phi_N | \psi_N \rangle|,
\label{eq:GGMdefn}
\end{equation}
where the maximization is taken over all \(N\)-party pure quantum states \(|\phi_N\rangle\) that are not genuinely multipartite entangled.
An \(N\)-party quantum state is said to be genuinely multiparty entangled if it is not a product state across any partition of the \(N\) parties 
that constitute the whole system.

 It turns out that above expression for GGM simplifies to \cite{amaderPRA, amaderbound}
\begin{equation}
{\cal E}(\psi_N) = 1 - \mbox{max} \{ \lambda^2_{A:B}| A\cup B = \{1,\ldots. N\}, A\cap B = \emptyset \},
\label{eq:GGMdefn1}
\end{equation}
where \(\lambda_{A:B}\) is the maximal Schmidt coefficient of \(|\psi_N\rangle\) in the bipartite split \(A:B\).

The measure possesses the usual properties of a genuine multiparty entanglement measure. In particular, it 
does not increase under local quantum operations and classical communication. 
Moreover, it is possible to compute GGM for an arbitrary multiparty pure quantum state of an arbitrary number of parties 
in arbitrary dimensions  \cite{amaderPRA, amaderbound} (cf. \cite{GM}).

 

\section{Classical Capacity of a Quantum Communication Channel}
\label{sec:cap}

The paradigmatic classical information transmission protocol over a quantum channel is quantum dense coding \cite{dc}.
In this paper, we will consider the dense coding protocol, by using the multiparty 
ground states of physically realizable quantum spin models, between several senders and a single receiver. 
For completeness, and to set the terminology, we begin with the case of 
%
 a single sender and a single receiver.

\subsection{Dense Coding: A Single Sender and a Single Receiver}

Let us suppose that an observer Alice ($A$), wants to send some classical information to another observer, Bob ($B$), who is in a distant laboratory,
and with whom Alice shares
a quantum state $\rho^{AB}$. 
Suppose that  $d_A$ ($d_B$) is the  dimension of the Hilbert space on which  Alice's (Bob's) part of the state 
$\rho^{AB}$ is defined.

Consider the situation where Alice wants to send the classical information \(i\), which is known to happen with probability \(p_i\), 
to Bob. The dense coding protocol now runs as follows. Depending on the information \(i\) that is to be sent, Alice performs the  
unitary transformation \(U_i\) (cf. \cite{MHPOVM}) on her side of the state $\rho^{AB}$,
 and sends her part of the quantum state to Bob over a noiseless quantum channel. Therefore, Bob now 
has the two-party ensemble \(\{p_i, \rho^{AB}_i\}\), where 
\(\rho^{AB}_i  =  U_i \otimes \mathbb{I}_B \rho^{AB} U_i^\dagger \otimes \mathbb{I}_B \), 
where \(\mathbb{I}_B\) is the identity operator on Bob's Hilbert space. 
Bob's aim is to gather as much information as possible, about the index \(i\),
by performing quantum mechanically allowed measurements on the two-particle ensemble that is in his 
possession now. This involves an optimization process, and it is possible to 
perform the optimization to obtain the dense coding capacity, of the state \(\rho^{AB}\), as 
\cite{dccapacity} 
\begin{equation}
C(\rho^{AB}) = \mbox{log}_2 d_A + S(\rho^B) - S(\rho^{AB}),
\label{eq:cap}
\end{equation}
where $S(\sigma) = -\mbox{tr} [\sigma \mbox{log}_2 \sigma]$ is the von Neumann entropy of the quantum state $\sigma$. 
In particular,  $S(\rho^{AB}) = -\mbox{tr} \left[\rho^{AB} \mbox{log}_2 \rho^{AB}\right]$ 
is the von Neumann entropy of the shared state $\rho^{AB}$.
And \(\rho^B = \mbox{tr}_A[\rho^{AB}]\) is
 a reduced density matrix of  \(\rho^{AB}\). Here the capacity is measured in ``bits''. 

To make the capacity values bounded by unity, independent of the 
dimensions in which we are working, we divide the actual capacity by the maximum (quantum mechanically) achievable capacity,
 $\mbox{log}_2 d_A+\mbox{log}_2 d_B$ bits, and we call this  form of the capacity as {\em normalized capacity}, which is then given by
\begin{equation}
\cal{C}(\rho^{AB}) = \frac{\mbox{log}_2 d_A + S(\rho^B) - S(\rho^{AB})}{\mbox{log}_2 d_A+\mbox{log}_2 d_B}.
\label{eq:capuni}
\end{equation}
Note that the normalized capacity is dimensionless.

\subsection{Dense Coding between Multiple Senders and a Single Receiver}

Let us now consider a dense coding protocol in which there are many senders, say Alice\(_1\), Alice\(_2\), \(\ldots\), Alice\(_M\) 
(denoted as 
$A_1,A_2, \ldots, A_M$), and a 
single receiver, say Bob (denoted as $B$).
The \(M\) Alices and Bob share the quantum state $\rho^{A_1,A_2,\ldots,A_M,B}$. 
Suppose that Alice\(_k\) $(k=1,2,\ldots,M)$ wants to send the classical information $i_k$ 
to Bob, 
where it is previously known that \(i_k\) appears with probability $p_{i_k}$. 

The dense coding protocol then runs as follows. To encode the message \(i_k\), Alice\(_k\) applies the 
unitary  $U_{i_k}$ on her part of the multiparty quantum state.
All the Alices then send their parts of the quantum state to Bob, over noiseless quantum channels, and so now Bob is in possession of the 
multiparty ensemble 
\be
\left\{\prod_{k=1}^M p_{i_k}, \otimes_{k=1}^M U_{i_k} \otimes \mathbb{I}_B         
\rho^{A_1,A_2,\ldots,A_M,B}
\otimes_{k=1}^M U_{i_k}^\dagger \otimes \mathbb{I}_B\right\},
\ee
 and his job is to find an optimal measurement strategy to obtain as much information as 
possible about the indices \(i_k\). 

It is possible to find an optimal strategy, and the resulting dense coding capacity 
is given by  \cite{dcmulti}
\begin{eqnarray}
C(\rho^{A_1,A_2,\ldots,A_M,B}) &=& \mbox{log}_2 d_{A_1} +\mbox{log}_2 d_{A_2}+\ldots+\mbox{log}_2 d_{A_M}\nonumber\\
&&\,\,\,\, + S(\rho^B) - S(\rho^{A_1,A_2,\ldots,A_M,B})
\label{eq:mulcap}
\end{eqnarray}
bits,
where $d_{A_k}$ is the dimension of the Hilbert space of Alice\(_k\)'s subsystem.
Here,  $S(\rho^{A_1,A_2,\ldots,A_M,B})$ is the von Neumann entropy of the state shared between the $M$ Alices and Bob.
Again, to normalize the capacity, we divide the actual capacity by 
$\mbox{log}_2 d_{A_1}+\mbox{log}_2 d_{A_2}+\ldots+\mbox{log}_2 d_{A_M}$ bits, to obtain the dimensionless normalized capacity 
as
\begin{widetext}
\begin{equation}
\cal{C}\left(\rho^{A_1,A_2,\ldots,A_M,B}\right) = \frac{\mbox{log}_2 d_{A_1} +\mbox{log}_2 d_{A_2}+\ldots+\mbox{log}_2 d_{A_M} + S\left(\rho^B\right) - 
S\left(\rho^{A_1,A_2,\ldots,A_M,B}\right)}{\mbox{log}_2 d_{A_1}+\mbox{log}_2 d_{A_2}+\ldots+\mbox{log}_2 d_{A_M}+\mbox{log}_2 d_B},
\label{eq:capnorm}
\end{equation}
\end{widetext}
where $\rho^B = \mbox{tr}_{A_1, A_2, \ldots, A_M} \left[\rho^{A_1,A_2,\ldots,A_M,B}\right]$, and \(d_B\) is the dimension of the 
Hilbert space of Bob's subsystem.

\section{Order-from-Disorder in Genuine Multipartite Entanglement and Channel Capacity}
\label{sec:ggmcap}

We will now investigate the behavior of genuine multipartite entanglement (as quantified by generalized geometric measure) and capacity of dense coding in multiport 
scenarios, in the ground states of different quantum \(XY\) spin models. In particular, we will be interested in the effect of the different types of 
disorder (as provided by the systems described by the Hamiltonians \(H_{SG}\), \(H^{RF}\), and \(H_{SG}^{RF}\)) 
on the multiparty entanglement measure and the multiport classical capacity, and compare them with the same quantities in the system described by the 
quantum \(XY\) Hamiltonian \(H\) in Eq. (\ref{XY}).

We consider the ground states of the models for the comparisons. This is dictated by the fact that while the GGM 
can be obtained efficiently for arbitrary \emph{pure} multiparty quantum states (of arbitrary number of parties and dimensions), it is much harder to find it for 
mixed quantum states. This feature of GGM is similar to that of 
most other measures of entanglement, where calculations for pure states is much easier than for mixed states. Exceptions include entanglement of formation (for 
two qubits) \cite{Wootters-EOF} and logarithmic negativity \cite{logneg}. 
However, both these examples are for two-party situations. The multiport channel capacity that we consider 
\emph{can} however be efficiently calculated for mixed states also. But since one of our main interests lie in the comparison of the behavior of 
GGM with that of the multiport capacity, with the introduction of disorder, we only work with the ground states of the models, and in regions in which 
there is no degeneracy in the ground state. We will have occasion to discuss more on the issue of degeneracy, later in the paper.


\subsection{Quantum $XY$ spin glass vs. quantum $XY$}
\label{ekta-batrish-1}

In this subsection, we compare the values of GGM and the multiport channel capacity in the quantum \(XY\) spin glass with those 
of the quantum \(XY\) model. 
The Hamiltonian for the (ordered) quantum $XY$ model is given in Eq. (\ref{XY}) and that for the (disordered)
 quantum $XY$ spin glass model Hamiltonian is given in Eq. (\ref{jihamil}). 

\begin{figure}[h]%
\resizebox{1.0\columnwidth}{!}{
\includegraphics{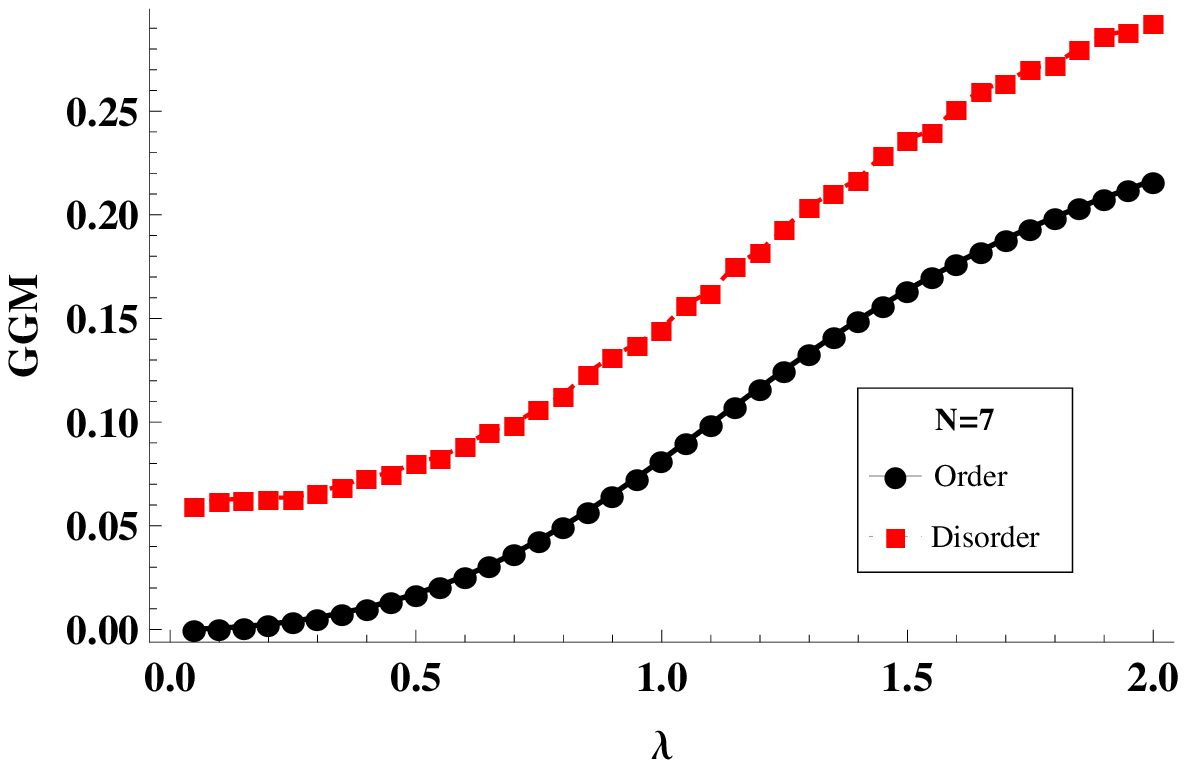}%
}
\resizebox{1.0\columnwidth}{!}{
\includegraphics{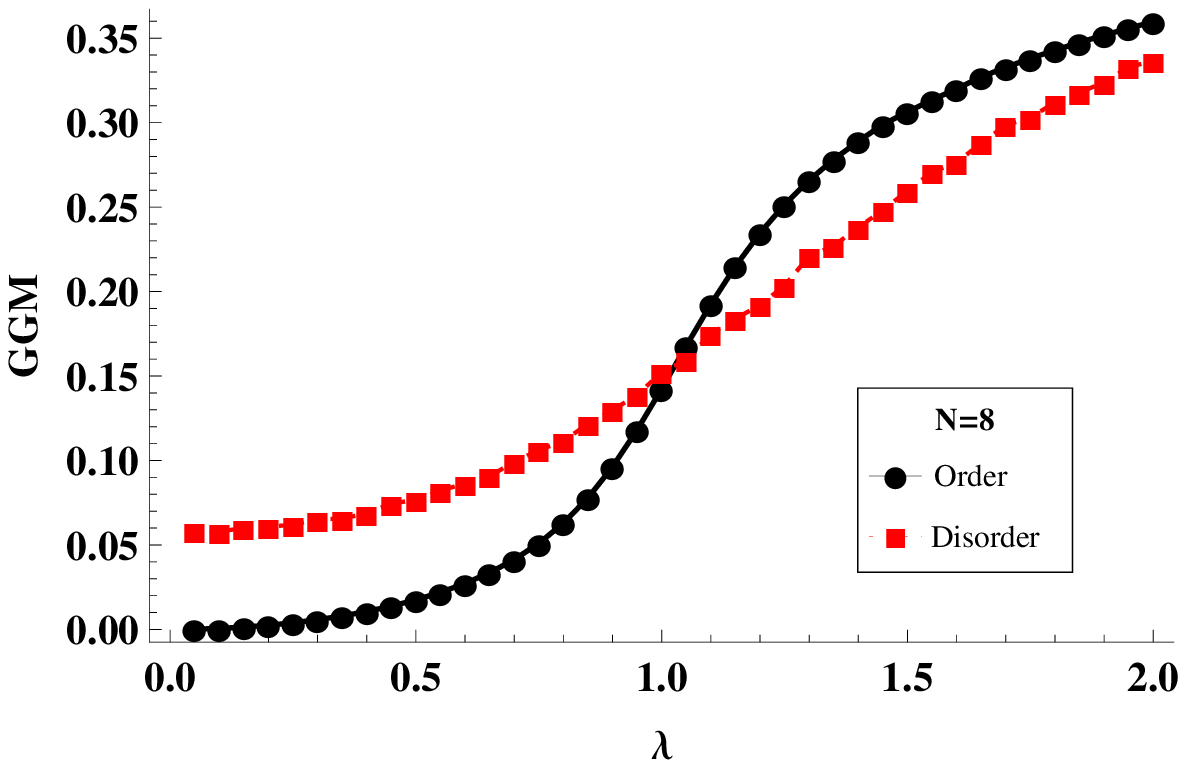}%
}
\caption{(Color online.) Order-from-disorder for a genuine multiparty entanglement in quantum \(XY\) spin glass. 
The upper plot is for 7 spins, while the lower one is for 8. In both the plots, GGM is plotted on the vertical axes.
The curves with black circles are for the ordered systems (with the Hamiltonian \(H\)), and in these cases, the GGM
are plotted against \(\lambda = J/h\) on the 
horizontal axes. 
The curves with red squares are for the disordered systems (with the Hamiltonian \(H_{SG}\)), and in these cases, the GGM are plotted against 
the mean values \(\lambda=J/h\) of the Gaussian distributed \(J_i/h\) (with mean \(J/h\) and unit standard deviation) on the horizontal axes.
The quenched averaging is performed over $5 \times 10^{3}$ disorder realizations.
Both axes represent dimensionless quantities in both the plots. The whole range of parameters considered offers disorder-induced enhancement of GGM for the case 
of an odd number of spins. For an even number of spins, there is a cross-over \(\lambda\), before which the disorder-induced enhancement occurs. 
%
See \cite{degeneracy-in-disordered-case} for the strategy assumed for the very small number degenerate ground states appearing.}
\label{fig:JN910ggm}%
\end{figure}

Let us first consider the calculations for GGM. For the ordered system, we find the GGM in the ground state of the system. For the disordered system, 
we find the GGM in the ground state for a particular realization of the disorder, and then average over the disorder. 
%
We perform calculations for upto 8 spins in a periodic chain. The results for odd and even number of spins are significantly different, and we display the plots here for 
\(N=7\) and for \(N=8\). See  Fig. \ref{fig:JN910ggm}. The plots for other odd and even number of spins are qualitatively similar, respectively to the presented plots. 
For the plots, we have chosen the anisotropy parameter (\(\gamma\)) in the Hamiltonians as 0.7. The plots for the other anisotropies are qualitatively similar.

 For the comparison, we find the GGM for the ordered system for a particular value of \(J=J{'}\) and \(h=h{'}\). We then find the quenched averaged GGM 
in the disordered system, where the transverse field is fixed at \(h{'}\), and where the \(J_i\)'s are all i.i.d. quenched 
random variables, with each \(J_i/h{'}\) following the Gaussian distribution with mean \(J{'}/h{'}\) and unit standard 
deviation. For each value of the mean, $J{'}/h{'}$, the quenched average is taken over \(5 \times 10^{3}\) realizations.
We have also performed numerical simulations for higher number of realizations, and have checked that the corresponding quenched physical 
quantities have already converged. 

For an odd number of spins, the quenched averaged GGM in the disordered system is always higher than the GGM in the corresponding ordered system. For an even number of 
spins, there is a ``cross-over value'', \(\lambda_c^{GGM}\), of \(\lambda= J/h\), before which 
the quenched averaged GGM in the disordered
system is larger than the GGM in the ordered system.

Therefore, there are distinct regimes where introduction of disorder enhances the amount of genuine multiparty entanglement present in the quantum system
\cite{qwe, qwe1}. The existence 
of entanglement in a quantum system, consisting of several subsystems, indicates the thermodynamic signature of high global order in the whole system while the local 
order in the subsystems is low \cite{Nielsen-Kempe}.
This disparity between global and local order is 
particularly pronounced in a maximally entangled state, e.g. the singlet state, where the global order is complete (the global state is pure) while
the local order is completely absent (the local states are completely depolarized). 
What we show is that this disparity between global and local order, in the sense of increasing genuine multipartite entanglement,
 can be enhanced by introducing disorder 
into the system. 
The interplay of global and local order to create entanglement is intrinsic to the multiparty quantum state 
under consideration. It is intriguing to see that agents of disorder that are external (in the sense that they can be externally manipulated on the system) 
can interact with the intrinsic order-disorder mechanism, to produce an increase of entanglement by increasing the disorder.

\begin{figure}[h]%
\resizebox{1.0\columnwidth}{!}{
\includegraphics{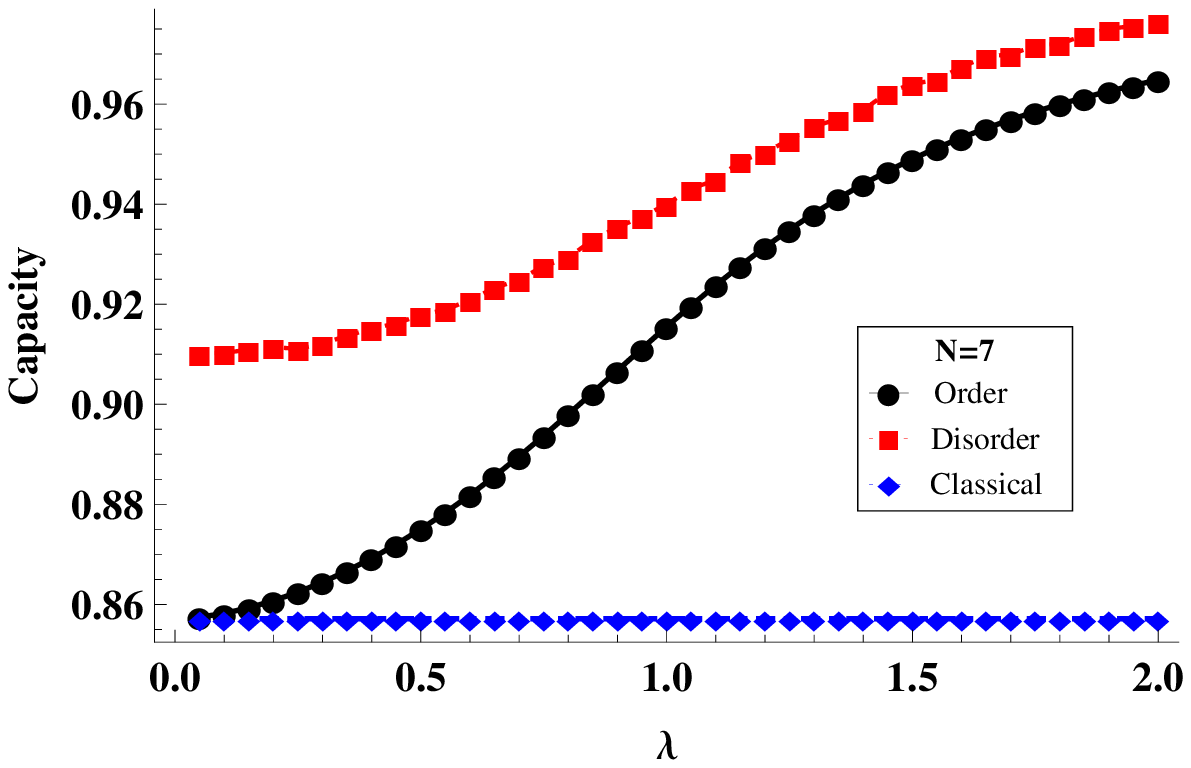}%
}
\resizebox{1.0\columnwidth}{!}{
\includegraphics{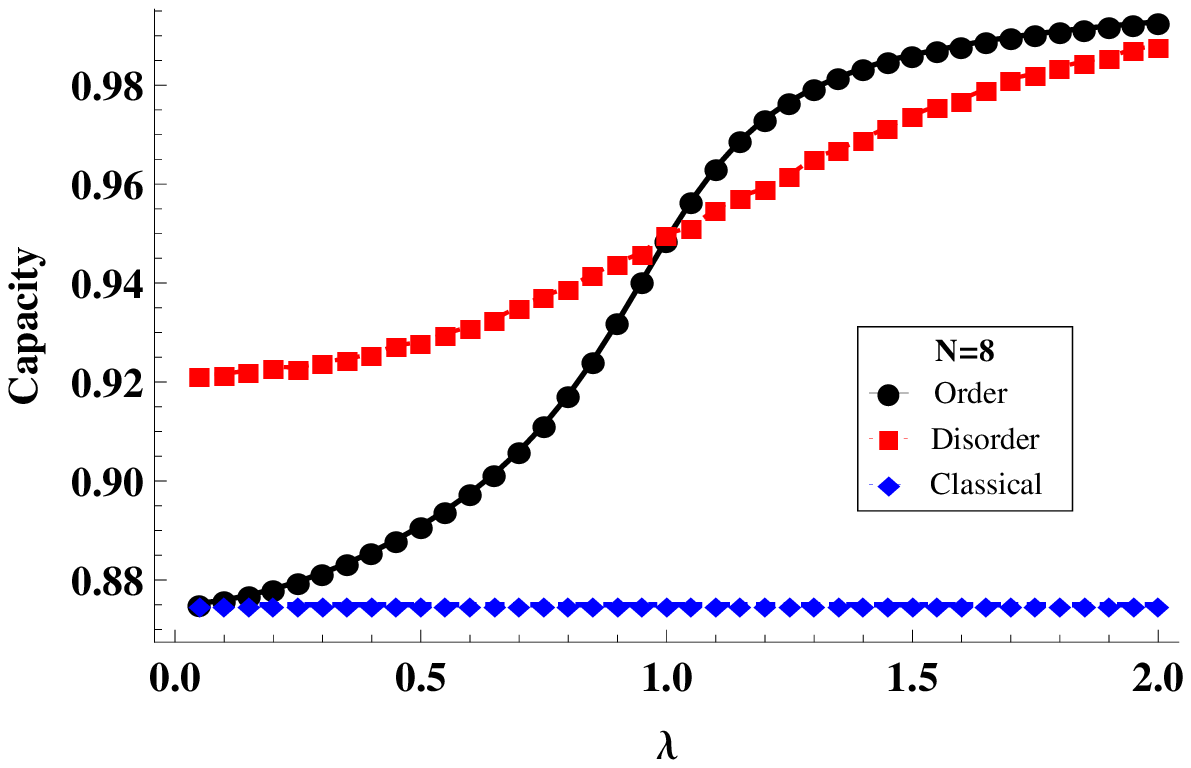}%
}
\caption{(Color online.) Order-from-disorder for a multiport classical capacity of a quantum channel in quantum \(XY\) spin glass. 
The upper and lower plots are for 7 and 8 spins respectively. 
In both the plots, the normalized capacity (dimensionless) are plotted on the vertical axes.
All other notations are the same as in Fig. \ref{fig:JN910ggm}, and the 
quenched averaging is performed over $5 \times 10^{3}$ disorder realizations.
Again, the whole range of parameters considered offers disorder-induced enhancement for the capacity for an odd number of spins, while there is a cross-over \(\lambda\)
before which the disorder-induced enhancement occurs for an even number of spins. The cross-over \(\lambda\) for the capacity is lower than that for the GGM.
The horizontal lines (with blue diamonds) correspond to the classical capacities that can be attained if there are no previously shared quantum states
 between the senders and the receivers.
}
\label{fig:JN910cap}%
\end{figure}

We  now move over to the comparison of the information carrying capacities of the same systems. The channel capacity that we consider will have multiple senders and a single
receiver. For the ground state of a system of \(N\) spins, we assume that \(N-1\) spins are in possession of \(N-1\) Alices (who act as senders) while the remaining spin
is in possession of Bob (who will act as the receiver of the information). Due to symmetry of the system, it is immaterial as to which spin is 
in possession of Bob, in the sense that the capacity will be independent of that choice.

Similarly as for the GGM, the multiport capacity behaves differently for odd and even number of spins. In Fig. \ref{fig:JN910cap}, we 
plot the normalized capacities for \(N=7\) and \(N=8\). The situation is qualitatively similar for other odd and even spins, respectively. 
Just like for GGM, the quenched averaged capacity in the disordered case is always better than the capacity in the ordered situation, if the total number of spins 
is odd, while for even \(N\), there is a cross-over \(\lambda_c^{cap}\), until which the capacity of the disordered case is better than that of the ordered Hamiltonian. 

Therefore, again there arises situations where introduction of disorder into the system increases the information carrying capacity of a quantum system.

Additionally, we find that it is necessary to have an order-from-disorder feature for  multiparticle entanglement,
to have the same in the 
multiport capacity. In other words, we observe that \(\lambda_c^{GGM} \geq \lambda_c^{cap}\) in all the cases considered. 

We term the disorder-induced enhancements for the genuine multiparty entanglement as well as the multiport capacity as ``order-from-disorder'', since 
 a higher efficiency (in the sense of increased genuine multiparty entanglement, which results in an increased multiport capacity, and possibly 
increased capacities for other quantum communication tasks) in the generated ground state is obtained by introducing disorder into the physical system under consideration.

\subsection{Quantum \(XY\): Random vs Non-random transverse fields}
\label{ekta-batrish-2}

\begin{figure}[h]%
\resizebox{1.0\columnwidth}{!}{
\includegraphics{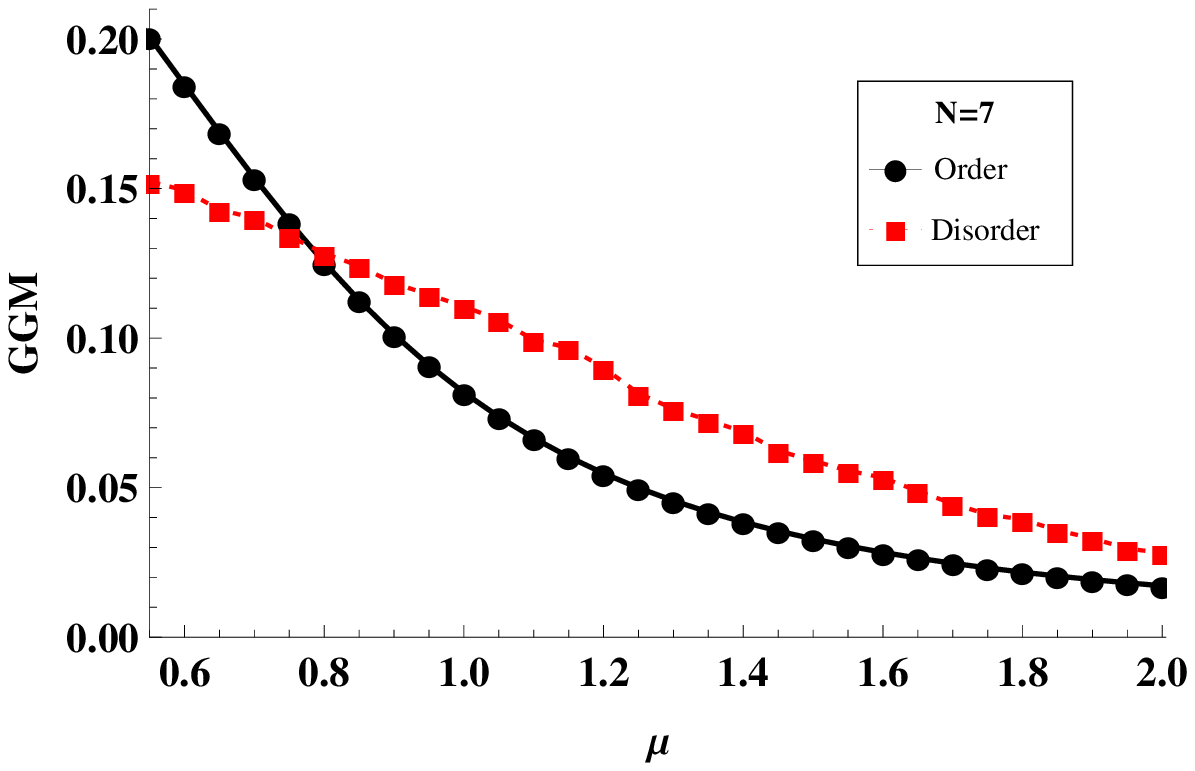}%
}
\resizebox{1.0\columnwidth}{!}{
\includegraphics{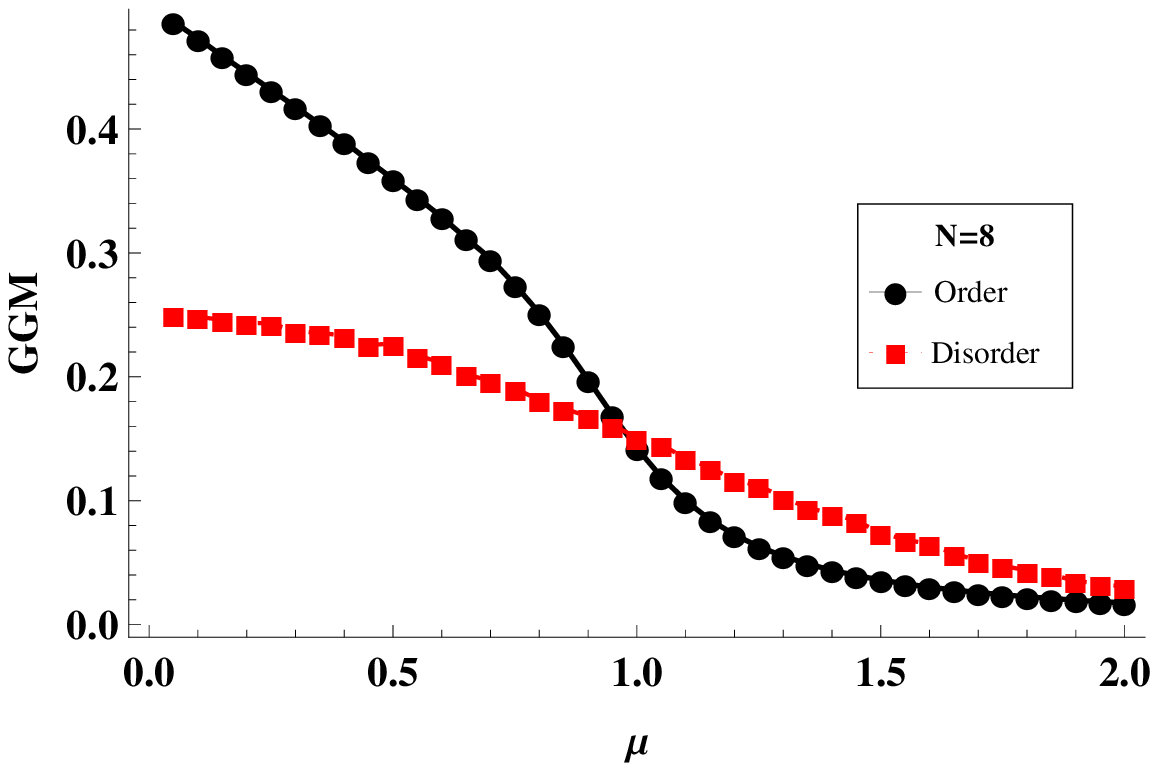}%
}
\caption{(Color online.) Order-from-disorder for GGM in quantum \(XY\) system with random transverse field. 
The upper plot is for \(N=7\), while the lower one is for \(N=8\). The vertical axes in both the plots represents the GGM. 
The curves with black circles are for the systems without disorder (represented by the Hamiltonian \(H\)), in which cases, the GGM are plotted against 
\(\mu = h/J\) on the horizontal axes. The curves with red squares are for the systems with disorder, and described by the Hamiltonian \(H^{RF}\), and 
in these cases, the horizontal axes represent 
the mean values \(\mu=h/J\) of the Gaussian distributed random variables \(h_i/J\) 
(with mean \(h/J\) and unit standard deviation). 
The quenched average is taken over $5 \times 10^{3}$ disorder realizations.
 All the axes represent dimensionless quantities. In contrast to the situation in Fig. \ref{fig:JN910ggm}, 
the cross-overs now appear for both odd and even \(N\).
}
\label{fig:hN910ggm}%
\end{figure}

For the ordered Hamiltonian, scanning over different values of the coupling constant is physically equivalent to scanning over
the transverse field. It may seem plausible that similarly, sweeping over different mean values of the coupling constant in the 
disordered Hamiltonian is equivalent to sweeping over mean values of a disordered transverse field.
However, we show below that introduction of a Gaussian disorder in the transverse field produces a qualitatively different behavior for the GGM as well as the multiport 
capacity, as compared to their behavior after introduction of a Gaussian disorder in the coupling constant. 

\begin{figure}[h]%
\resizebox{1.0\columnwidth}{!}{
\includegraphics{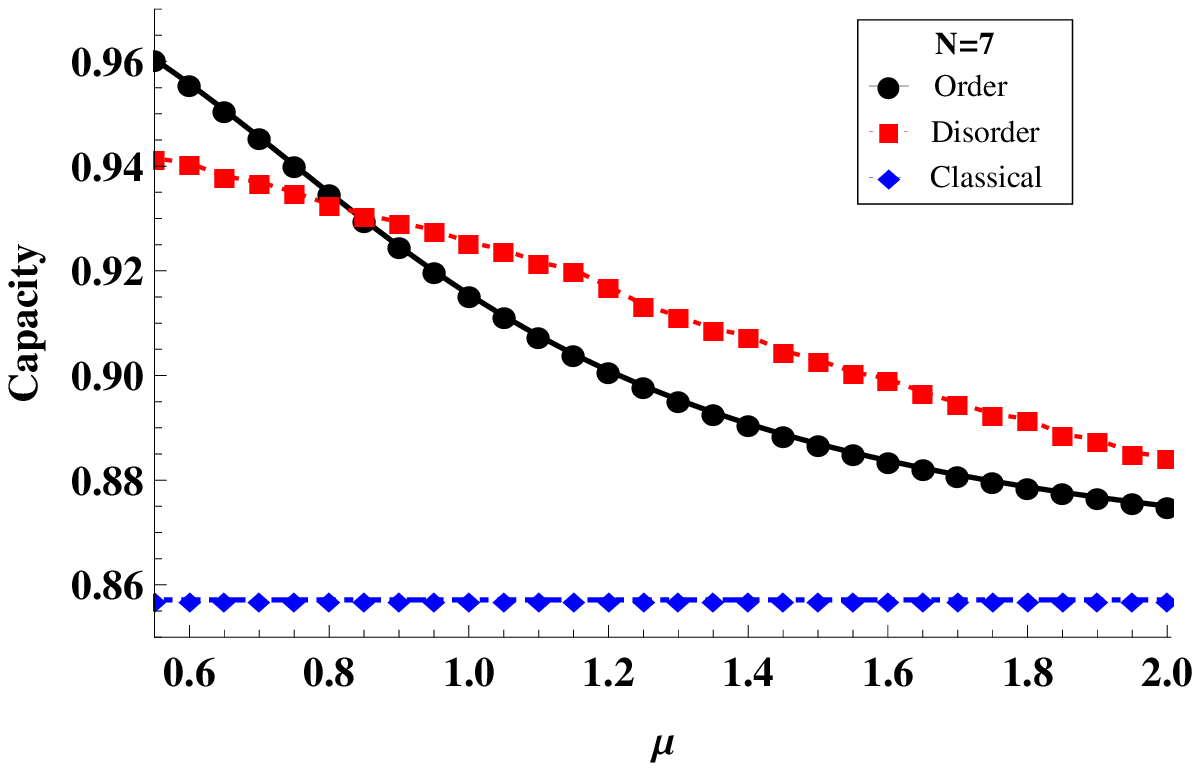}%
}
\resizebox{1.0\columnwidth}{!}{
\includegraphics{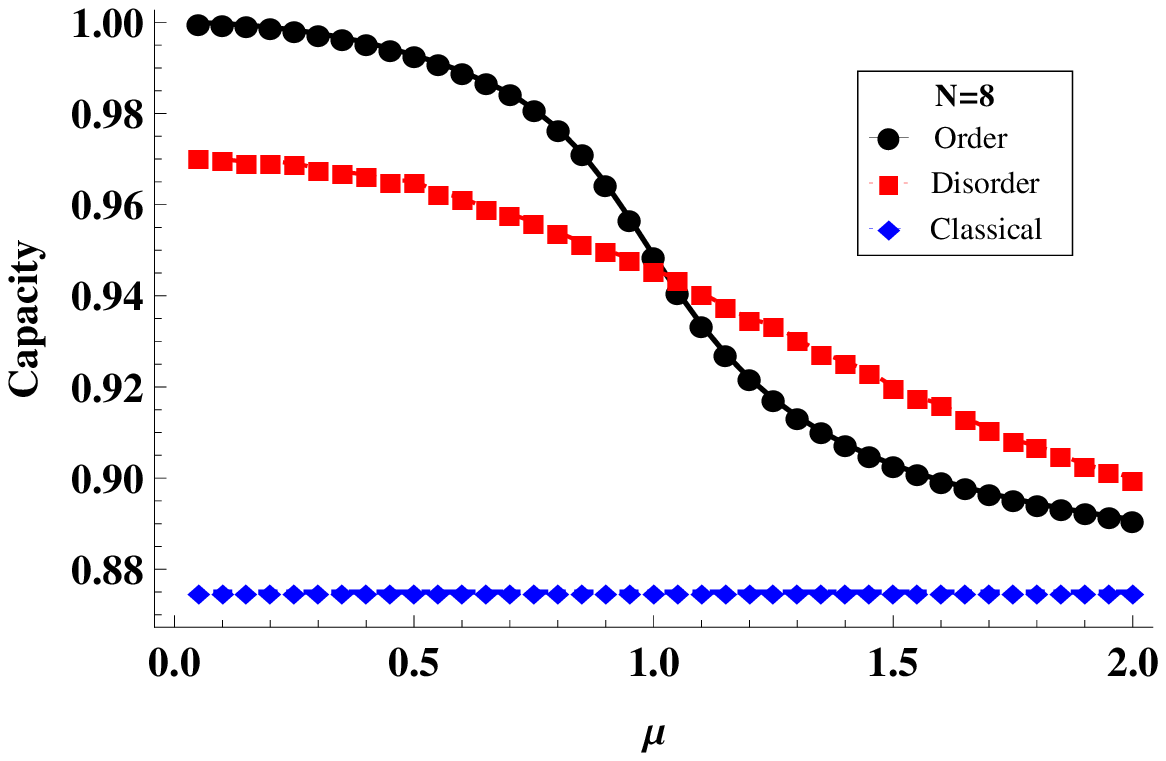}%
}
\caption{(Color online.) Order-from-disorder for the multiport capacity in quantum \(XY\) system with random transverse field. 
The vertical axes in both the plots represent the normalized capacity (dimensionless). 
All other notations remain the same as in Fig. \ref{fig:hN910ggm}, and  $5 \times 10^{3}$ disorder realizations are used for 
the quenched averaging. There are cross-overs again, and the cross-over \(\mu\)'s for the 
capacity are higher than those for GGM. The horizontal lines (with blue diamonds) have the same meaning as in Fig. \ref{fig:JN910cap}.
}
\label{fig:hN910cap}%
\end{figure}

As in the preceding subsection, we consider the ground state of the ordered Hamiltonian (Eq. (\ref{XY})) for  given \(J=J{'}\) and \(h=h{'}\), and find the 
GGM and the multiport classical capacity. We then compare them with those obtained from the ground state of the disordered Hamiltonian in Eq. (\ref{hihamil}), 
for  \(J=J{'}\) and for i.i.d. quenched random variables \(h_i\), with each \(h_i/J\) distributed as \(N(h{'}/J{'},1)\),
i.e., with mean \(h{'}/J{'}\) and unit standard deviation.
%
And again, in the disordered case, the averaging over the 
disorder is performed after the physical quantity has been calculated for randomly obtained values of the disorder parameter. See Figs. 
\ref{fig:hN910ggm} and \ref{fig:hN910cap}.

In contrast to the situation when disorder is introduced in the coupling constants (as discussed in the preceding subsection), 
there are now cross-overs for both odd and even total number of spins. 
That is, for both odd and even total number of spins, 
there are cross-over values, \(\mu_c^{GGM}\), of \(\mu=h/J\), before which the
 quenched averaged physical parameters are lower than the corresponding physical parameter in the ordered Hamiltonian, and 
after which the situation is the opposite. 

There does however appear a difference between the odd and even cases, but it is of a separate kind. Precisely, the ordered Hamiltonian for an 
odd number of spins has a degenerate ground state for low \(\mu\), and as said before, we will therefore disregard this range of 
\(\mu\). The ground state is however nondegenerate for the whole  range of \(\mu\) for an even number of spins. This feature for odd \(N\)  was also obviously 
present in the considerations of the preceding subsection, for high \(\lambda\), and were also disregarded there.

Just like when we introduced disorder in the coupling constants, again we have obtained distinct regimes where introduction of 
disorder in the transverse field enhances the genuine multipartite entanglement as well as the multiport channel capacity of the quantum system. And 
again, a disorder-enhancement in GGM is necessary for a similar feature in the capacity.



\subsection{Quantum \(XY\) spin glass with random transverse field vs. quantum $XY$}
\label{ekta-batrish-3}

\begin{figure}[h]%
\resizebox{1.0\columnwidth}{!}{
\includegraphics{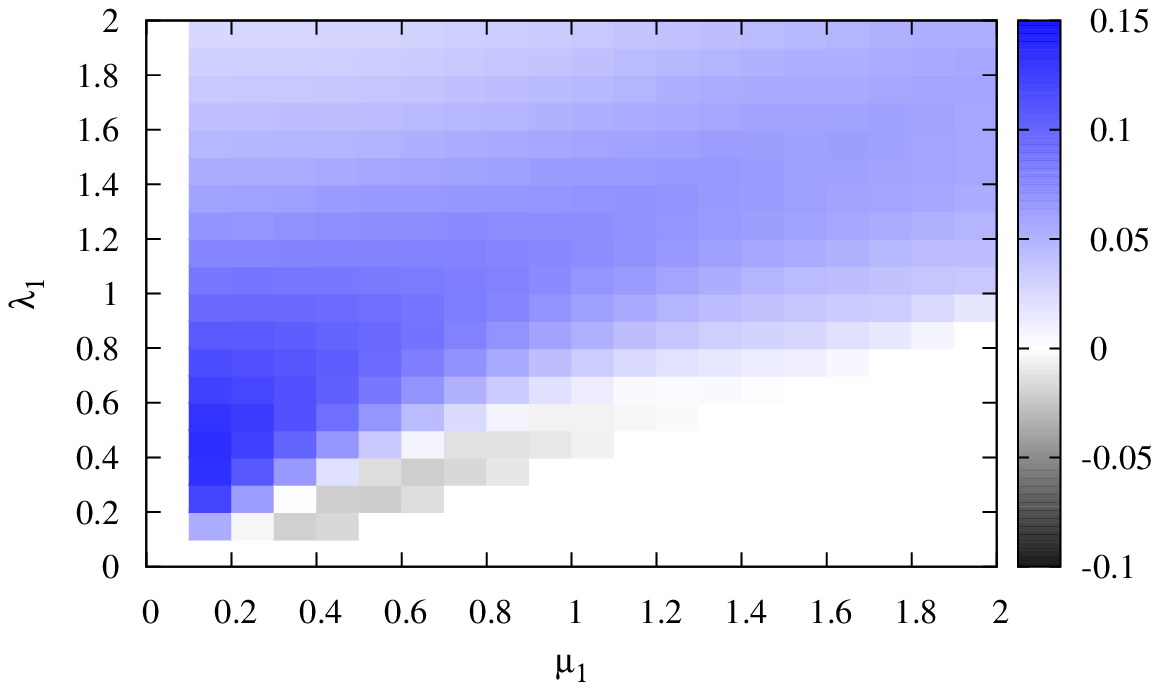}%
}
\resizebox{1.0\columnwidth}{!}{
\includegraphics{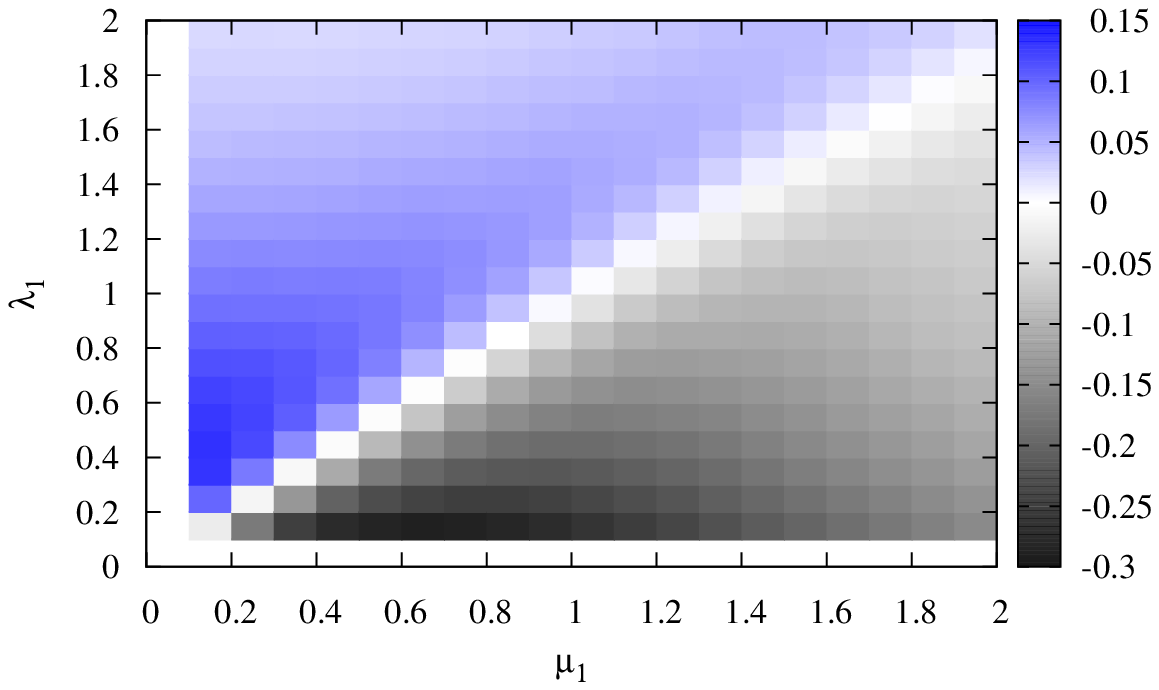}%
}
\caption{(Color online.) Order-from-disorder for genuine multiparty entanglement in quantum \(XY\) spin glass with random transverse field.
The upper plot is for \(N=7\) and the lower one is for \(N=8\). The vertical axes represent $\lambda_1=J/\kappa\), while the horizontal axes represent $\mu_1=h/\kappa\), with 
\(\kappa >0\). 
 The plotted quantity is the difference (disorder case \(-\) order case) between the quenched averaged GGM for 
the system described by \(H_{SG}^{RF}\) (with i.i.d. random variables \(J_i\) and \(h_i\), where the \(J_i/\kappa\) are distributed as \(N(J/\kappa,1)\) and 
the \(h_i/\kappa\) are distributed as \(N(h/\kappa,1)\)) and the GGM for the system described by \(H\).
%
Again, the quenched averaging is performed over $5 \times 10^{3}$ disorder realizations.
See text for further details.
The plotted quantity as well as the quantities represented by the axes are dimensionless.}
\label{fig:jhN78ggm}%
\end{figure}

We have already investigated the situations where 
disorder is introduced into the quantum system through separate channels -- via coupling constants and via 
transverse magnetic fields. Introduced separately, both have given rise to regimes where 
disorder enhances the amount of a genuine multiparty entanglement as well as the amount of a multiport channel capacity. 
The question that we want to raise now is whether the two types of disorder, if introduced \emph{together}, will still allow for regimes of 
disorder-enhanced physical quantities. This is not a priori obvious as the effects of disorder-enhancement due to the two types of disorder may not 
work in tandem, and may, in principle, completely wash out the disorder-enhancement phenomena. This however is not the case, as we show now. 



\begin{figure}[h]%
\resizebox{1.0\columnwidth}{!}{
\includegraphics{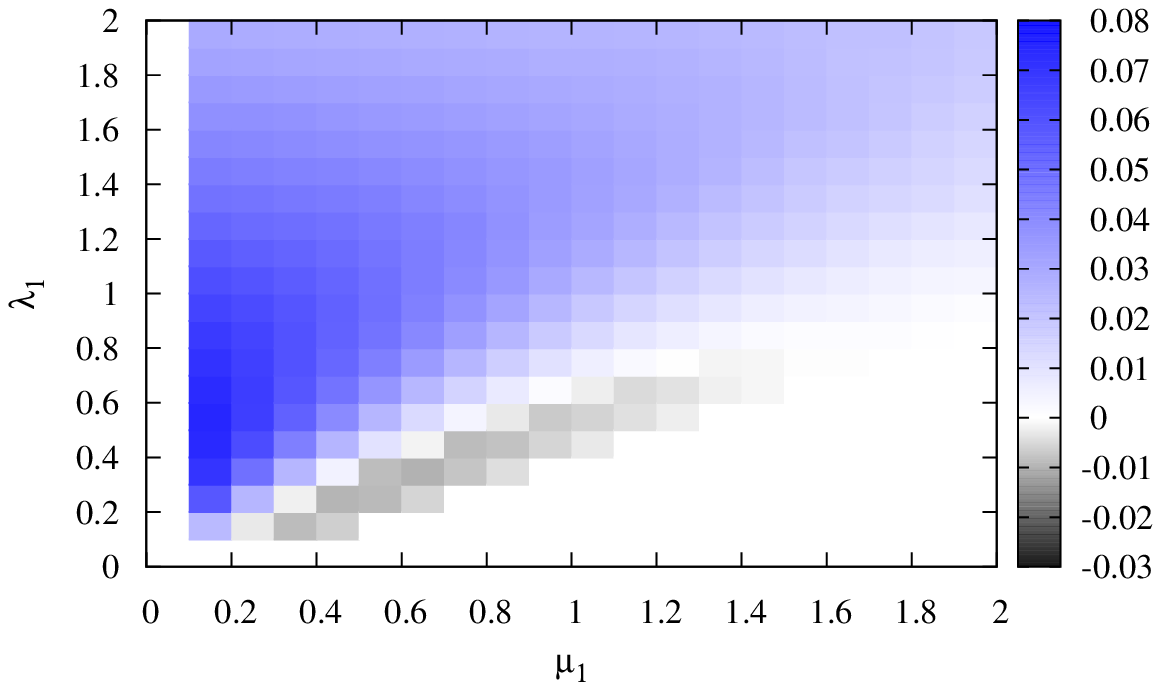}%
}
\resizebox{1.0\columnwidth}{!}{
\includegraphics{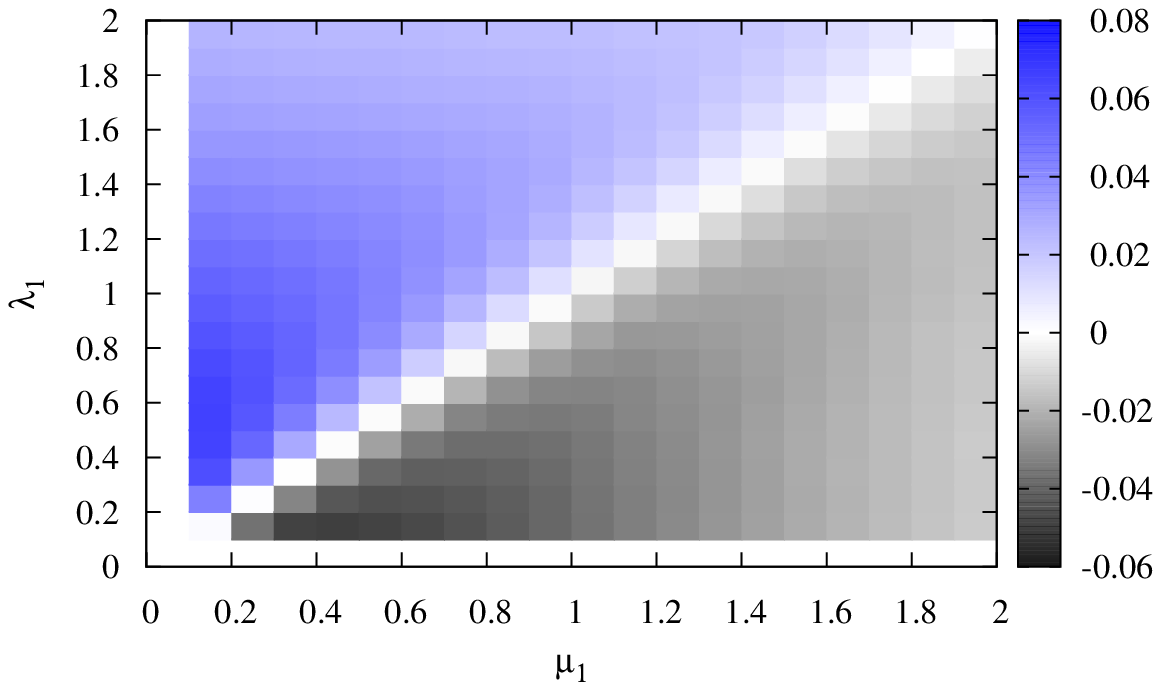}%
}
\caption{(Color online.) Order-from-disorder for the multiport capacity in quantum \(XY\) spin glass with random transverse field.
All notations are the same as those in Fig. \ref{fig:jhN78ggm}, except 
that 
the plotted quantity is the difference between the multiport capacities in the two situations considered. 
Again, the quenched averaging is performed over $5 \times 10^{3}$ disorder realizations.
Also see text for further details. 
}
\label{fig:jhN78cap}%
\end{figure}

Consider a particular realization of the ordered Hamiltonian in Eq. (\ref{XY}) for \(J=J{'}\) and \(h=h{'}\). 
Correspondingly, let us consider the Hamiltonian \(H_{SG}^{RF}\), for i.i.d. quenched random variables \(J_i\) and
\(h_i\), where the \(J_i/\kappa\) are distributed as \(N(J{'}/\kappa,1)\) and the \(h_i/\kappa\) are distributed as \(N(h{'}/\kappa,1)\). 
Note that \(\kappa\) is positive and has the unit of energy. 
The quenched average is taken over $5 \times 10^{3}$ disorder realizations.
We compare the GGM and the capacity in the ground states of the two Hamiltonians. 
The results are plotted in Figs. \ref{fig:jhN78ggm} 
and \ref{fig:jhN78cap}. 
For the plots, we find the difference (disorder case \(-\) order case) of the values of a particular physical quantity (GGM or capacity) as obtained in the 
ordered case and in the corresponding disordered case (after quenching).
As before, the situations are different for odd and even \(N\). 
The results are stated below separately for the two cases. For specificity, we consider the cases \(N=7\) and \(N=8\), and \(\gamma = 0.7\), in the plots. 
The 
situation is similar for other odd and even total number of spins, and for other anisotropies. 

We begin with the case where the total number of spins is odd. There are three qualitatively different regions. See the upper plots in 
Figs. 
\ref{fig:jhN78ggm} 
and \ref{fig:jhN78cap}. 
\begin{itemize}
 \item \textbf{Regions I.} In the upper plot in Fig. \ref{fig:jhN78ggm}, this is the whole \((J,h)\) region shown, 
except for roughly a quadrant of an ellipse with corners 
approximately at (0.2,0), (2,1), and (2,0). [Online, this region is blue in color.] In this region, the quenched averaged GGM in the disordered Hamiltonian 
is higher than the GGM in the corresponding ordered system. A similar, but slightly smaller, region is in the upper plot of Fig. \ref{fig:jhN78cap}, where 
the quenched averaged capacity is greater than that in the ordered Hamiltonian. [Online, this region is also blue in color.] This latter region is
contained in the former region, again implying that a disorder-enhancement in GGM is a pre-requisite for disorder-enhancement in the capacity.

\item \textbf{Regions II.} These are tiny portions of the \((J,h)\) region of both the plots, just below Region I, in the bottom left corners of 
the plots.
[Online, these regions are black to grey in color.]
In these regions, there is no disorder-enhancement. 

\item \textbf{Regions III.} Finally, in the remaining portions of the \((J,h)\) regions plotted, the ordered Hamiltonians have degenerate ground states, and 
and are thereby disregarded. These are the white regions in the bottom right of the plots. They should not be confused with the white 
boundary between Regions I and Regions II, which indicates an equality between the quenched averaged physical quantity and that obtained from the ordered Hamiltonian.  
\end{itemize}

Going over to the case where the total number of spins is even, we find that the the picture is simpler here, as there is no degeneracy observed in the 
region under study, and consequently 
there are no Regions III here. See the lower plots  in 
Figs. 
\ref{fig:jhN78ggm} 
and \ref{fig:jhN78cap}. The other two regions are however distinctly present, and placed approximately on two sides of the antidiagonals. 
\begin{itemize}
\item  \textbf{Regions I.} In case of the GGM, which is considered in the lower plot of Fig. 
\ref{fig:jhN78ggm}, Region I is above the antidiagonal \(J=h\). In this region, the quenched averaged GGM is higher than the ordinary variety. [Online, this region is 
blue in color.] The status of the multiport capacity is considered in the lower plot of Fig. \ref{fig:jhN78cap}, where the Region I is similar, but slightly smaller, to 
the Region I in the lower plot of Fig. \ref{fig:jhN78ggm}. Again, the latter region is contained in the former region. 

\item  \textbf{Regions II.} This is the complement of the Regions I, and indicates that the quenched averaged values are lower than the ordinary ones. 
 
\end{itemize}

\begin{figure}[h]%
\resizebox{1.0\columnwidth}{!}{
\includegraphics{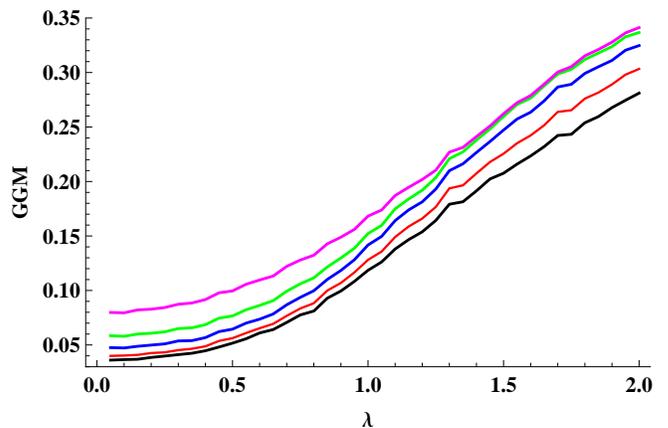}%
}
\caption{(Color online.) Behavior of quenched averaged GGM for different anisotropy parameters in the spin glass system. We consider a system of 8 spins. 
The vertical axis represents the quenched averaged GGM. 
The different curves are for different values of the anisotropic parameter \(\gamma\). From bottom to top, they are respectively for
\(\gamma =0.1\) (black), 0.3 (red), 0.5 (blue), 0.7 (green), and 1.0 (magenta). The GGM is plotted for the ground state of the spin glass system
described by the Hamiltonian \(H_{SG}\), and is plotted against the mean values \(\lambda = J/h\) of the Gaussian quenched random variables \(J_i/h\). The standard deviations 
of the random variables are all unity. 
The quenched averaging of GGM is taken over $5 \times 10^{3}$ disorder configurations.
Both the axes represent dimensionless quantities. 
}
\label{fig:ggmgamma}%
\end{figure}

\begin{figure}[h]%
\resizebox{1.0\columnwidth}{!}{
\includegraphics{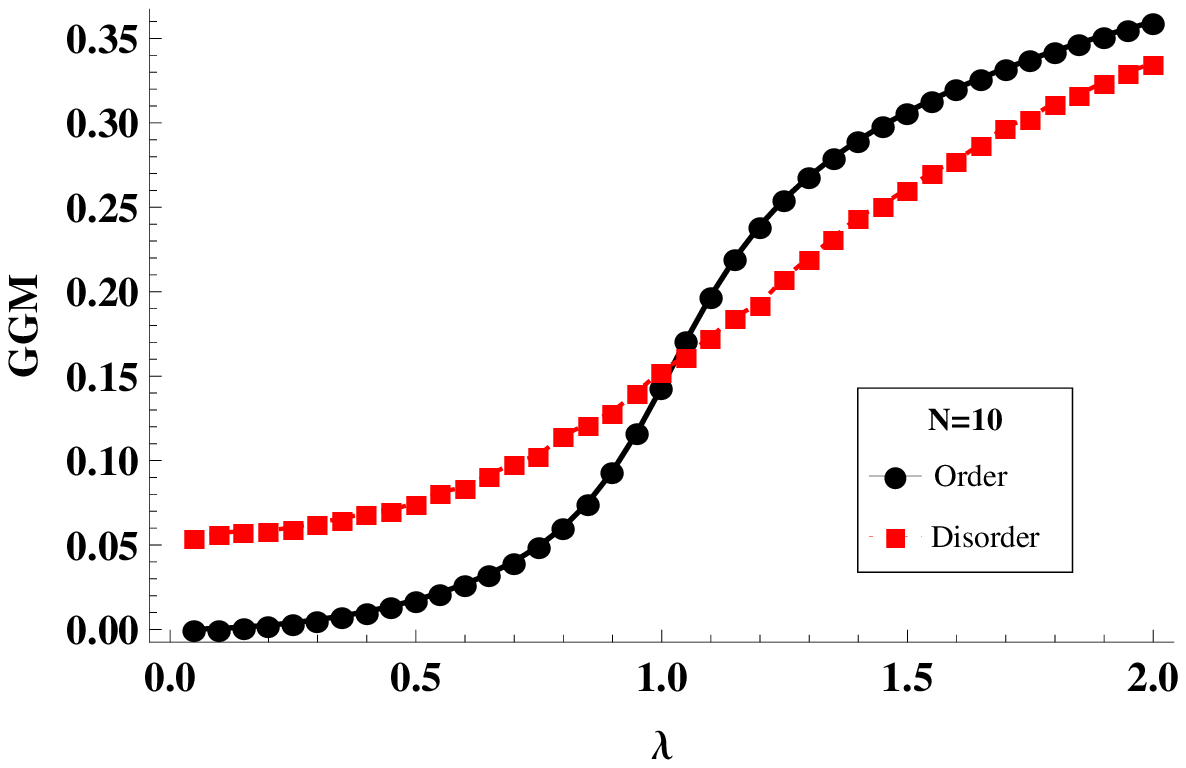}%
}
\resizebox{1.0\columnwidth}{!}{
\includegraphics{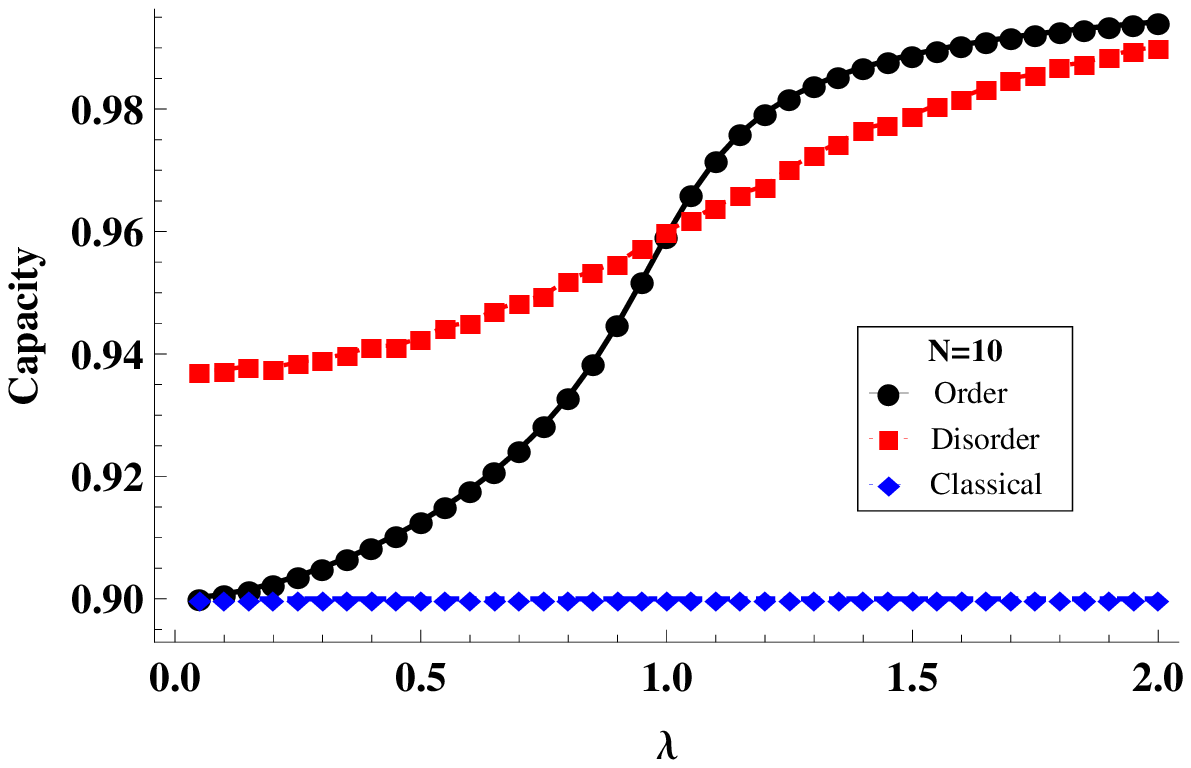}%
}
\caption{(Color online.) Order-from-disorder for the genuine multiparty entanglement and capacity in quantum \(XY\) spin glass with 
transverse field. Here we consider \(N=10\) spins. All notations are the same as in Fig. 2. The quenched averaging is performed over $5 \times 10^{3}$ disorder realizations.
}
\label{10spinsevidence}%
\end{figure}

\section{Discussion}
\label{conclusion}


Summarizing, we have found that the ground states of several quenched disordered quantum spin systems, 
 in the presence of certain ranges of the disorder parameters, have higher genuine multipartite entanglement than in their  counterparts 
in systems without disorder.
In almost the same range of parameters,
the ground states of disordered models turn out to be  better carriers of classical information than their equivalents in systems 
without disorder.
The models considered include the quantum anisotropic $XY$ spin glass with and without a random  transverse magnetic field.
%
The calculations are performed by exact diagonalizations, and subsequent quenched averaging, whenever required. 

The results shown in the plots displayed in the paper
are given, for definiteness, for a specific value of the anisotropy parameter \(\gamma\). However, we have performed the calculations for a large range of \(\gamma\), and 
the general qualitative behavior is similar for all \(\gamma\). To exemplify the genericity of the qualitative behavior obtained from the previous plots in the paper,
let us present a figure (Fig. \ref{fig:ggmgamma}) 
where we plot the quenched averaged genuine multipartite entanglement for different values of \(\gamma\), in the spin glass 
Hamiltonian \(H_{SG}\).

One of the aims of the study was to investigate 
the relation between genuine multiparty entanglement and the capacity of classical information transfer 
in these specific systems. 
The results displayed until now are all for systems of eight spins, and this is dictated by the 
approximate number of spins which can be experimentally accessed coherently in quantum systems. 
However, such boundaries are steadily being extended and improved \cite{ionexp, Monroe, Schmidtkaler, photonexp, Zeilinger, 14qubit, 4photon, 56photon}. 
In this light, it is important to mention here that the results obtained in this paper are resilient with respect to moderate changes of the total number of 
spins. As an evidence for that, in Fig. \ref{10spinsevidence}, we show that disorder-induced advantage seen until now for \(N=8\) is valid also 
for \(N=10\). 

We term the phenomenon of increased genuine multiparty entanglement and increased multiport capacity subsequent upon the introduction of 
disorder into the system as order-from-disorder. Experimental and theoretical work in quantum information processing in the last two decades or so have 
gradually convinced us that the phenomenon of entanglement, observed in a variety of physical systems, is not a fragile quantity. We hope that this 
work will add to that increasing belief.

\begin{acknowledgments}

We acknowledge computations performed at the cluster computing facility in HRI (http://cluster.hri.res.in/).
\end{acknowledgments}


\begin{thebibliography}{99}
\bibitem{degeneracy-in-disordered-case} In all the disordered cases considered, there has appeared only a few instances of degeneracy in the ground state of 
the corresponding Hamiltonian. Indeed, for almost all values of the means of the disorder parameters, there has been no degeneracy. In the small number of cases, 
where a given value of the common mean of the i.i.d. random variables  allows for some set of values of the i.i.d. random variables, such that the corresponding 
Hamiltonian has a degenerate ground state, the number of such sets is never more than 0.5\%, and in most cases, it is 
substantially lower. These sets are then removed from the family of sets that is fed into the 
quenched averaging. This is the strategy that is followed  throughout the paper.
Due to the low frequency of the appearance of the sets with a degenerate ground state, and since both GGM and the multiport capacity are 
bounded functions, the final result is qualitatively independent of these degenerate sets.
\end{thebibliography}
\end{document}